# PI LOOP RESONANCE CONTROL FOR DARK PHOTON EXPERIMENT AT 2 K USING A 2.6 GHz SRF CAVITY

C. Contreras-Martinez[†], B. Giaccone, O. Melnychuk, A. Netepenko, Y. Pischalnikov, S. Posen, and V. Yakovlev, Fermilab, Batavia, Illinois, USA

*Abstract*

Two 2.6 GHz SRF cavities are being used for a dark photon search at the vertical test stand (VTS) in FNAL, for the second phase of the Dark SRF experiment. During testing at 2 K the cavities experience frequency detuning caused by microphonics and slow frequency drifts. The experiment requires that the two cavities have the same frequency within the cavity's bandwidth. These two cavities are equipped with frequency tuners consisting of three piezo actuators. The piezo actuators are used for fine-fast frequency tuning. A proportional-integral (PI) loop utilizing the three piezos on the emitter was used to stabilize the cavity frequency and match the receiver cavity frequency. The results from this implementation will be discussed. The integration time was also calculated via simulation.

## INTRODUCTION

A dark photon is a hypothetical particle that weakly couples to ordinary matter and is an extension of the standard model (SM) of particle physics [1]. The experiment discussed in this paper is one of several which are known as "light-shining-through-wall" experiments [2]. The first phase of the Dark SRF experiment at FNAL was done using two single-cell 1.3 GHz cavities [3]. The next step consists of using two high $Q_o$ single-cell 2.6 GHz SRF cavities, one of which will be called the emitter cavity and the other the receiver cavity. They will be first tested at VTS and later be moved to the dilution refrigerator to improve sensitivity. The emitter will be powered on creating an electromagnetic field ($TM_{010}$) inside the cavity, the field from this emitter cavity acts as a source of dark photons that can be emitted outside the cavity. The dark photons can penetrate the receiver cavity and be converted to SM photons [3] which have the same frequency as the emitter. This conversion will result in a signal on the receiver. A schematic of this process is shown in Fig. 1. The field excited in the receiver is proportional to the field in the emitter [2,3].

Resonant enhancement of the receiver signal is achieved when the frequency of the SM photons, that result from the conversion of the dark photons from the emitter cavity, matches the frequency of the receiver cavity. The bandwidth of the cavities is given in Table 1. Note that the receiver cavity has a small bandwidth of 0.56 Hz which puts a constraint on the peak detuning that the cavity should experience. Additionally, the cavity is extremely sensitive to deformation with a sensitivity of 10 kHz/µm. Both cavities initially consist of three Physik Instrumente PICMA actuators. The piezos were later changed to stainless-steel rods on the receiver cavity only. The total range of the tuner is ~ 176 kHz (240 V range), more detail on the tuner for these cavities can be found in [4]. During testing microphonics and slow frequency drifts were observed.

During this experimental run, the dark photon search was not conducted. Instead, the feasibility of maintaining the cavity frequencies close to each other was explored. The emitter cavity was powered on and the frequency shift of both cavities was recorded. Once the frequency drift of both cavities was characterized a PI loop was implemented only on the emitter cavity to mitigate the cavity frequency mismatch. An estimate of the integration time is given based on the PI loop resonance control of the emitter cavity.

Table 1: Figures of merit of both cavities, the bandwidth is calculated from the loaded $Q_L$

| Cavity | R/Q [Ω] | Bandwidth [Hz] | $Q_L$ $\times 10^8$ |
|---|---|---|---|
| Emitter | 104.7 | 5.84 | 4.42 |
| Receiver | 104.7 | 0.56 | 46.4 |

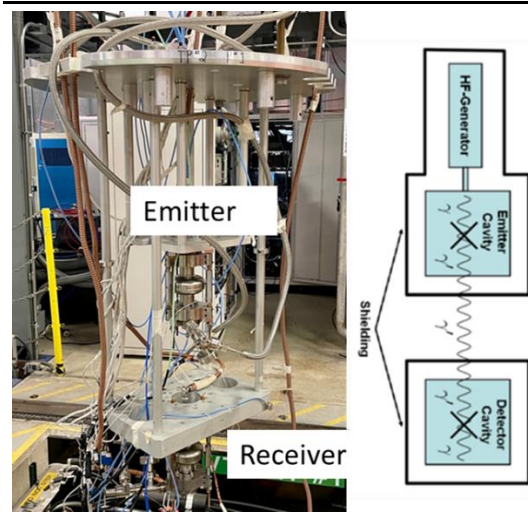

Figure 1: Left picture shows the setup of the experiment. The right picture shows a schematic of the process of dark photon production and detection [5].

This material is based upon work supported by the U.S. Department of Energy, Office of Science, National Quantum Information Science Research Centers, Superconducting Quantum Materials and Systems Center (SQMS) under contract number DE-AC02-07CH11359

† ccontrer@fnal.gov

# CAVITY FREQUENCY DRIFT

From previous experience operating in liquid helium, it is known that the cavity will experience microphonics and slow frequency drifts. The frequency drift was characterized with a test conducted in VTS, where both cavities are placed inside a Dewar and submerged with liquid helium at 2 K. The next phase of this experiment will be to move the 2.6 GHz setup to a dilution refrigerator where the dark photon search will be conducted. The cavity frequency was recorded with two different methods, one using the VTS RF system which utilizes a frequency counter. The other method used a network analyzer (NA) model number Keysight E5080B. When comparing both cavities at the same time a NA is used on the receiver cavity.

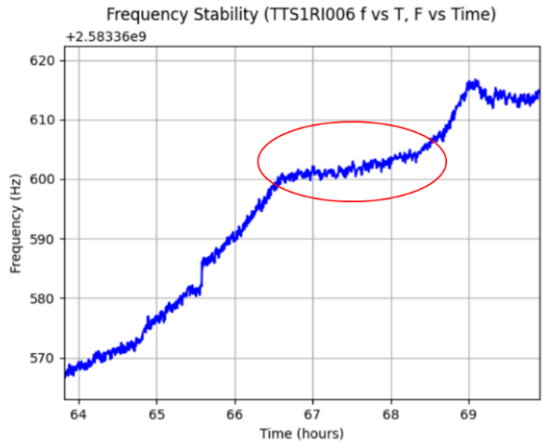

Figure 2: Frequency stability of the receiver cavity using a network analyzer. The cavity frequency drift can be linear or can plateau (Piezos are shorted on the receiver cavity).

Fig. 2 shows the frequency shift of the receiver cavity, the rate of change is ~ 10 Hz/hr, both cavities exhibit this behavior when excited by a low field with the NA. The cavity frequency was recorded with a NA. The frequency drift can shift from linear rise or decrease. There are also periods when the cavity frequency almost plateaus (circled on the plot). The frequency drift when the emitter cavity is excited to a field of 15 MV/m is shown in Fig. 3. The frequency drift, in this case, is 657 Hz/hr. This frequency was recorded with the VTS system that uses a frequency counter. This frequency shift for the emitter while large can easily be compensated with a PI loop. A PI loop is not feasible on the receiver cavity since exciting the cavity's $TM_{010}$ mode would hide the dark photon signal. Therefore, to maintain the same frequency between the two cavities a search for the culprit of the frequency drift observed in Fig. 2 was conducted. There are four possible causes for the slow frequency drift observed. One possibility is that the piezo can be driving the drift either through noise or piezo creep. The second is due to the helium bath pressure, and the third is due to small temperature drifts. The temperature sensitivity at 2 K is -20 kHz/K. The fourth reason is the stability (fluctuation) of the driving frequency for the cavity. This could come from the NA output port or the input port.

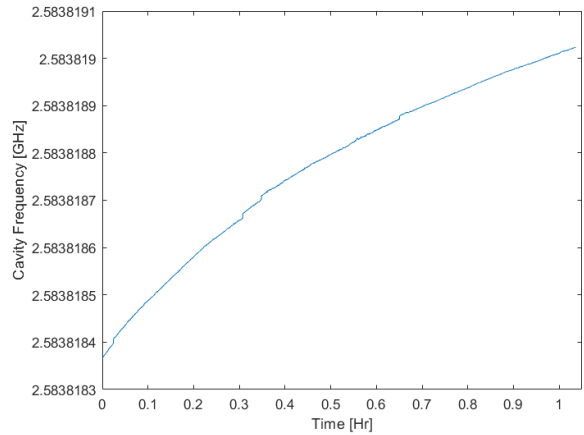

Figure 3: Frequency stability of the emitter cavity with a gradient of 15 MV/m and piezo DC voltage of 108 V.

The piezo material (PZT) has local domains each with a dipole moment. When an electric field is applied the domains that are aligned with the field either expand or contract depending on the direction of the field. The piezo creep arises from the lag of the orientation of the dipole moment domains with the electric field. This results in a slow expansion (or contraction) until a steady state is reached [6]. To see a 10 Hz frequency shift the piezos would need to expand 1 nm. The results from Fig. 2 are for the case when the piezos are shorted. Therefore, it is unlikely that the shifts are due to the piezo creep since no voltage is applied or the noise through the amplifier. If piezo creep was causing the frequency drift it will only move in one direction and the plateau (see Fig. 3). This is not the behavior observed in Fig. 2. In the trial where the PI loop was implemented (see Fig. 6) the three piezos were replaced on the receiver cavity with stainless-steel rods and invars screws to remove the uncertainty about whether the piezos were causing the drift. The results from Fig. 6 show a similar frequency drift as in the circled region in Fig. 2. These results show that the piezos did not contribute to the drift with the piezo creep or noise.

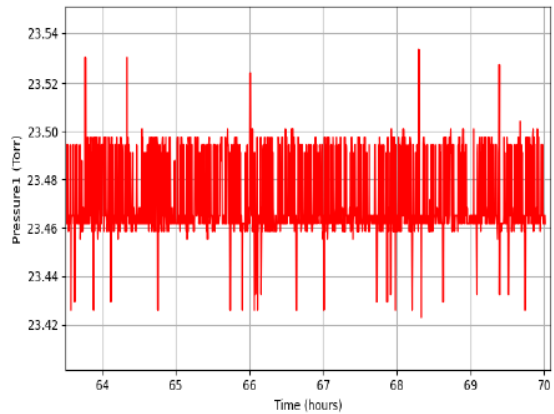

Figure 4: Pressure measured in the Dewar; the location of the pressure sensor is located above the cavities.

The pressure inside the Dewar was recorded during the same period as in Fig. 2, the results are shown in Fig. 4.

The cavities have a df/dp ~-25 Hz/Torr. The pressure fluctuations observed in the figure are on the order of 0.04 Torr for most of the data. Note that these fluctuations could also be due to noise from the sensor. These variations will result in a frequency shift of 1 Hz. Some pressure peaks can reach a shift of 0.1 Torr which can result in a 2.5 Hz frequency shift. The pressure time series doesn't have a linear trend like the cavity frequency shift. The pressure sensor is located above the cavity. This then demonstrates that it is not well correlated with the frequency shift. More measurements are needed to exclude liquid pressure from contributing to the drift. For the temperature drift to affect the cavity frequency a shift of 500 µK must be observed. The RTD sensors are mounted directly on the surface of the cavity near the equator. The RTD results show no correlation with the drift.

The frequency stability of the NA is tested by using a Rohde Schwarz SMA100B signal generator as a stable frequency source. The signal was connected to the input of the NA. The frequency was monitored for 50 minutes and no change was observed from the initial value of 2583 MHz. The RF generator frequency was then shifted by 1 Hz to verify that the system had not stalled. These results demonstrate that the input port on the NA is stable.

Based on these results there is no clear source on what is causing the cavity frequency drift. More work needs to be done on monitoring the pressure inside the Dewar and the temperature of the cavities which are the most likely culprits.

## PI LOOP RESONANCE CONTROL

The dark photon search needs the cavity frequencies to be within the bandwidth of the receiver cavity which is 0.56 Hz. A PI loop was used on the emitter to achieve this. Ideally, both cavities must have the PI algorithm but the receiver cavity cannot be excited at the same frequency as the emitter cavity to avoid hiding the dark photon signal. A higher order mode at ~5.2 GHz ($TM_{020}$) was used to monitor the cavity frequency on the receiver but the higher order mode and $TM_{010}$ mode were not well correlated. To correlate both modes the cavity was excited by both modes simultaneously. The $TM_{010}$ frequency shift was recorded with the NA analyzer and the $TM_{020}$ mode detuning was recorded with the frequency counter. The frequency detuning observed in the $TM_{020}$ mode was twice as large compared to the $TM_{010}$ mode.

Note that when a PI loop is used the piezo creep is not a concern since the algorithm will adjust the voltage to mitigate this frequency shift. In this iteration the receiver cavity did not have piezos installed but instead stainless-steel rods with the same diameter as the piezos but at half the length were used. To compensate for half the length invar screws were used to prevent large shrinkage. More details can be found in [4].

The emitter cavity frequency and power are controlled by the VTS RF system. The cavity frequency is monitored with a frequency counter model Agilent 53132A. The frequency is read via GPIB. The sampling is limited to 10 Hz or less due to frequency counter averaging and the system not being able to do a higher value. During the measurement, the sampling frequency was set to 4 Hz. For control of the frequency drift observed in Fig. 2 the sampling rate is enough since the frequency shift is 2 mHz/s. The PI loop algorithm is implemented using LabView on a computer. The algorithm computes the voltage that the piezo must be set to maintain the emitter cavity frequency close to the frequency of the receiver cavity. The computer is connected to an analog voltage output driven by the calculation of the PI algorithm, this signal is connected to the piezo amplifier. The number of samples that are sent out of the analog output is 4 samples/s. A schematic of this setup is shown in Fig. 5.

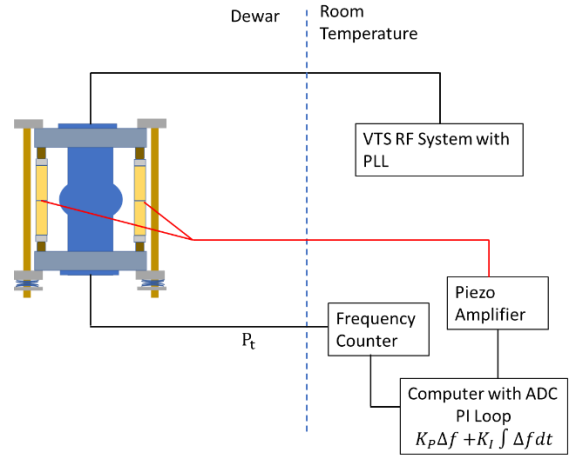

Figure 5: Setup of the PI loop for the emitter cavity.

The PI algorithm computes the difference between the emitter cavity frequency and the set frequency ($\Delta f$). The set frequency is the value of the receiver cavity frequency at t=0 s when the algorithm starts. The emitter cavity is powered on using the VTS RF system at 15 MV/m. The frequency of the receiver cavity is measured with the NA. The piezo voltage is adjusted so that the emitter cavity frequency matches the receiver cavity. After they are matched the PI algorithm starts.

Fig. 6 shows the results of the implementation of the PI loop on the emitter cavity. During this run only the integrable component of the algorithm was used. The trapezoidal rule is used to integrate the $\Delta f$ signal. An integral gain ($K_p$) of 2 was used to stabilize the frequency. Using this gain resulted in a rate shift of -0.1 Hz/hr. For future tests a larger gain can be used for further stabilization. During this period the piezo voltage decreased by 0.4288 V as shown in Fig. 7. This voltage decrease is equivalent to an increase of frequency of 314 Hz during this time without the PI loop and with a set piezo voltage of 115 V DC. The reason why the cavity frequency would have gone that high could be due to the creep (in this case mitigated by the PI loop), the heating of the cavity since it was powered on, and Lorentz force detuning (LFD).

While the overall frequency rate of change on the emitter is -0.1 Hz/hr, which is good, microphonics with peak detuning of 2 Hz is also observed in Fig. 6. What causes the microphonics cannot be resolved with the frequency coun-ter measurement. Additionally, the receiver cavity frequency has a drift of 1.7 Hz/hr. Note that the frequency change of the receiver was recorded using the NA. During a 2.2 Hr acquisition, the receiver cavity had a frequency drift of 3.6 Hz which is roughly seven times its bandwidth. The effects of this frequency mismatch will be discussed in the next section.

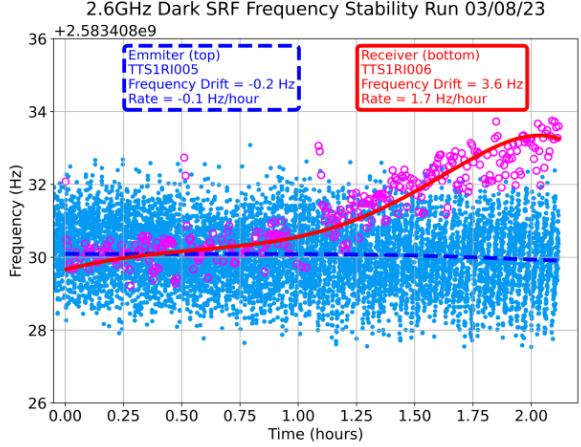

Figure 6: Frequency drift of the emitter cavity is shown in blue and for the receiver it is shown in red. A frequency counter is used to record the frequency on the emitter and NA for the receiver.

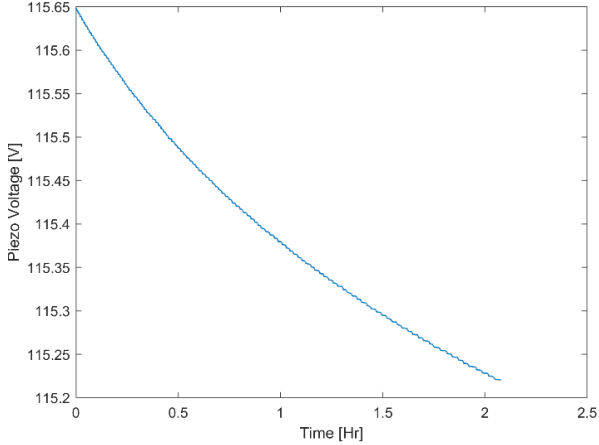

Figure 7: Voltage of the piezos on the emitter cavity with PI algorithm.

## INTEGRATION TIME SIMULATION

The previous section showed how the cavity frequencies were not matched within the receiver cavity bandwidth even when using a PI loop on the emitter cavity. The effects of this mismatch caused by slow frequency drifts are calculated using a simulation to estimate the field inside the receiver cavity. The cavity field behavior was modeled by using an LCR circuit, note that the same equation can be arrived at by using the cavity field and expanding in the cavity eigenmodes as shown in [2,7]. The second-order differential equation can be reduced to the first since only slow variations are considered [8]. The equation to model the emitter cavity voltage is given by

$$\frac{dV_E}{dt} + (\omega_{1/2} - i\Delta\omega_E)V_E = R_L\omega_{1/2}(I_{RF}) \quad (1)$$

Where $V_E$ is the emitter voltage, $\omega_{1/2}$ is the cavity half bandwidth, $\Delta\omega_E$ is the detuning of the emitter cavity, $R_L$ is the loaded shunt impedance, and $I_{RF}$ is the current that drives the circuit, in this case from the RF power. The solution when the RF power is constant is given by

$$V_E = \frac{R_L\omega_{1/2}I_{RF}}{\omega_{1/2} - i\Delta\omega_E}\left[1 - e^{-(\omega_{1/2} - i\Delta\omega_E)t}\right] \quad (2)$$

The field excited in the receiver cavity is proportional to the one in the emitter cavity [2]. Eq. 1 is used to model the voltage of the receiver except the driving term is

$$I_E \propto \frac{\frac{\omega_1}{2}}{\frac{\omega_1}{2} - i\Delta\omega_E(t)}\left[1 - e^{-\left(\frac{\omega_1}{2} - i\Delta\omega_E(t)\right)t}\right] \quad (3)$$

which is proportional to the voltage of the emitter cavity. Eq. 1 is solved using Runge-Kutta 4 on MATLAB to model the voltage of the receiver cavity with different frequency drifts.

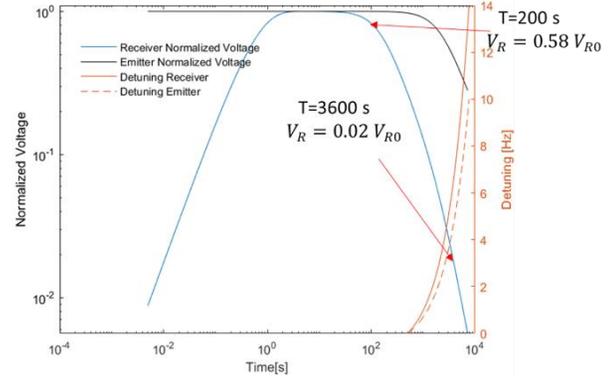

Figure 8: The simulations are done with no PI loop and linear cavity frequency drifts. For the emitter frequency change rate is 5Hz/hr and for the receiver it is 2 Hz/hr.

The detuning of the emitter is given by $\Delta\omega_E = \delta\omega_{slow,E}$ and for the receiver cavity $\Delta\omega_R = \Delta\omega_E + \delta\omega_{slow,R}$. The slow frequency shift for the emitter is given by $\delta\omega_{slow,E}$ and for the receiver it is $\delta\omega_{slow,R}$. Two different frequency drifts are used since this is what was observed in the data in the previous sections. Note that the frequency shift of the receiver $\Delta\omega_R$ also includes the frequency shift of the emitter $\Delta\omega_E$. For the simulations at t = 0 s the frequency of the emitter and receiver are matched. The RF generator of the emitter cavity is set to the frequency of the receiver. Two scenarios are modeled, the first scenario for when the emitter has no PI loop and the second is when the PI loop is implemented on the emitter.

The results of the simulation with no PI loop are shown in Fig. 8. The simulations run for a simulated 2 hours. The figure shows the receiver and emitter voltage which are normalized to the maximum voltage of each cavity. The detuning rate is modeled as a linear trend with 5 Hz/hr for the emitter and 2 Hz/hr for the receiver. With no PI loop on the emitter the voltage in the receiver drops to 58 % after 200 s and to 2 % at 1 hour. The result for the simulation with a PI loop on the emitter is shown in Fig. 9. The rate of change for the frequency is given in the figure. The voltage in the receiver drops to 94 % after 200 s and 15 % at 1 hour. The simulations discussed only consider the slow frequency drifts. Microphonics are present on both cavities, if the microphonics vibrations are in phase for both cavities it won't be an issue since both cavities will detune at the same time. If they are out of phase then this will create an additional frequency mismatch. Additional simulations are needed to take this into account.

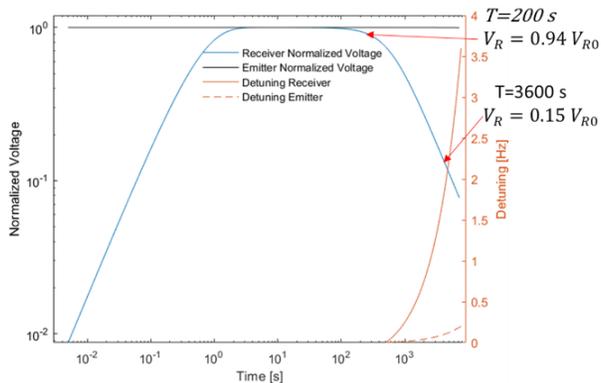

Figure 9: PI loop on the emitter, Emitter frequency change rate is 0.1 Hz/hr and for the receiver it is 1.7 Hz/hr

## CONCLUSION

A mock test for dark photon search was conducted in VTS. The cavity frequency measurement shows that there is a slow frequency drift that creates a mismatch frequency between the cavities. The source of this slow drift has not been identified, more work on finding it is needed. The receiver cavity frequency has small bandwidth and the slow frequency drift creates a frequency mismatch between the two cavities. A PI loop was implemented on the emitter cavity. Using PI resonance control on the emitter reduces the slow drift from 657 Hz/hr to 0.1 Hz/hr. This improvement in frequency stabilization improved the frequency matching capability which will greatly help increase the dark photon search sensitivity. Note that the 2.6 GHz dark photon search is planned to be conducted in a dilution refrigerator where the drift might not be present.

A simulation to estimate the integration time was done. The results show that even with a PI loop on the emitter after 1 hour the field of the receiver drops to 15 % of its initial value. A constraint of integration time is calculated to be about ~200 s when only using the PI loop on the emitter.